%% file: paper.tex
\documentclass[12pt]{article}

\usepackage[margin=1.0in]{geometry}

\usepackage{amsmath,amsthm,amssymb}
\usepackage{fancyhdr}
\usepackage{csquotes}
\usepackage{multicol}
\usepackage{url}
\usepackage{xurl}
\usepackage[backend=bibtex]{biblatex}
\PassOptionsToPackage{hyphens}{url}
\begin{document}

\title{A study on the Morris Worm}
\author{Akshay Jajoo
}
\maketitle
\begin{multicols}{2}

\input abstract
\input acknowledgement
\input overview
\input lphm
\input wormAndPatches
\input lessonsLearnt
\input changes
\input costAna
\input summary
\input appendix

\printbibliography
\end{multicols}
\end{document}

%% file: abstract.tex
\section*{Abstract}
\nocite{*}
The Morris worm \cite{techreportByPeter,techreportBySpaf} was one of
the first worms spread via the internet. It was spread on
November 2, 1988, and changed how
computer security was viewed by computer professionals as well as
general public \cite{giacByLarry}. Since its inception the Morris
worm has been studied extensively from the security point of view
\cite{techreportBySpaf,fifteenYear,techreportByDonn, techreportByPeter}
and is still a point of interest \cite{washingtonPost,
kasperSky}. \\ 
This paper summarizes the effects,
impacts, and lessons learned from the episode. There are other copies of this paper present at \cite{jajooMwormPurdue, jajooMwormGithub} and \cite{jajooMwormResearchGate}. However, I recommend using the arXiv version only.

%% file: acknowledgement.tex
\section*{Acknowledgement}
This paper was started as part of CS 52600 Information Security class at Purdue by prof. Eugene H. Spafford.
Thanks to prof. Spafford guiding the paper. I would also like to thank my friend Ben Harsha for helping with 
proof reading the paper. Thanks to Lovepreet Singh for all his help.

%% file: overview.tex
\section{Overview}
On November 2, 1988 at around 6 PM a Cornell university grad-student, Robert
Morris \cite{morrisHomepage, morrisWikipediapage}, launched a computer worm
(see \S\ref{sec:appendix} for explanation of worm) with the intention(he claims) of
mapping the Internet. The worm was self-replicating and self-propagating and
took advantage of exploits in the Unix services, described in detail in
\S\ref{sec:lphm:vuln}.  Though Robert claims that the worm was only intended for
educational purposes, the worm disrupted the whole Internet
\cite{washingtonPost} ( more in \S\ref{sec:lphm:inno}). 
Being first of its kind, the incident was very interesting for computer
scientist. Many researchers at Berkely, MIT and Purdue studied the worm and
uncovered how it works and released some fixes and patches within a day.
Technical details of the patches are discussed in \S\ref{sec:overviewAndPatches}.
The worm also impacted the way computer security was perceived, right from the
formation of CERT to people being more cautious and thoughtful about security.
Some people even term the episode as the big bang of cybersecurity \cite{njvcBlog}.
\S\ref{sec:learning} and \S\ref{sec:changes} talk more about learning and changes 
which followed the event. At that time
there were roughly 60,000 machines on the Internet, and it is estimated that the
episode resulted in a loss of \$98 million. Another estimate says that by
the time incident was isolated around 5-10 \% machines on the Internet were
victimized \cite{giacByLarry}. How drastic could such an incident be with many
companies and much of our day to day life activities solely relying on computers?
\S\ref{sec:costAna} discusses this in detail. \S\ref{sec:summary} is summary
and \S\ref{sec:appendix} talks about some other but related interesting facts 
and terminologies.

%% file: lphm.tex
\section{Loopholes and Misfeatures}\label{sec:lphm} 
As soon as the worm was noticed individual researchers, University committees, and government agencies all started analyzing the vulnerabilities exploited by the worm.
\cite{cornellCommission, techreportBySpaf, techreportByUSGeneralAccountingOffice, techreportByDonn}.
This section discusses what went wrong and what were the loopholes? 
Which exploits of BSD UNIX the worm took advantage of to reach its target?
Did Morris made some innocent mistakes in writing the replicating routine? Or was the worm
intentionally designed to spread like forest fire?

\subsection{Vulnerabilities Exploited} \label{sec:lphm:vuln} Along with many
other flaws, the episode exposed a few specific loopholes in services provided by
BSD-derived versions of UNIX. Researchers, engineers and system administrators
identified and reported \cite{techreportBySpaf,techreportByDonn,mitRochlis}
these bugs within 2 days of the outbreak. The following vulnerabilities were exploited.
\begin{enumerate} 
\item \textit{Sendmail} -- \textit{Sendmail} is a
mailer program used to route emails on the Internet.
The program was capable of running in various modes, one of them was
being a background daemon. In this mode the program used to \textit{listen}
on a TCP connection for an incoming mail. \textit{Sendmail} allowed mail
to be delivered to a process (the background daemon) instead of the mailbox
files, which was used for purpose like setting up automatic vacation responses.
Also, while fixing some other security bugs in the \textit{sendmail} \cite{techreportByDonn}
accidentally a new misfeature was introduced in it. The new bug was that if the
\textit{sendmail} is compiled with \textit{DEBUG} flag on and if the sender of
a mail asks the daemon to go in debug mode by sending a \textit{debug} command,
then the \textit{sendmail} allowed the sender to send a sequence of
commands instead of a recipients address. A combination of these two features was exploited in the worm.  
\item \textit{fingerd} -- The \textit{fingerd}
the utility was used to obtain general user information like name, username and
current login status of other users on the system. Like \textit{sendmail}, \textit{fingerd} 
also ran as a background daemon. The worm exploited \textit{fingerd} by overrunning the
buffer the process used as its input.\\
The source of the bug was the
\textit{gets} routine in the standard C library. A call to \textit{gets} writes
input to a buffer and the flaw was that the function implementing \textit{gets} assumed that the buffer passed
to it is large enough for the input to be written. This fault was not
only in \textit{gets}, but also, exists in other input-output routines like
\textit{scanf} and \textit{fscanf}.
\item \textit{rsh} and \textit{rexec} -- \textit{rsh} and \textit{rexec}
are services which offer remote command interpreters over a network. For
authentication purposes \textit{Rsh} used permission files and a privileged
source port whereas \textit{Rexec} used passwords. The worm exploited the fact
that there is a high possibility that a password for a local user for an
account on a remote machine will be same as its local password. Another
likelihood was that a remote account for a user will have \textit{rsh}
permission files for the local account of the user. The worm exploited above
two ideas to penetrate into remote machines.  Use of \textit{rsh} was very
simple -- just look for an account on a remote machine for a user who is running
the worm locally. Use of \textit{rexec} was not that simpler. To use 
\textit{rexec} for penetrating the worm used the local password to do a remote login.
So for doing this the worm had to crack local password first. 
Following subsection discusses the password cracking.
\item \textit{Passwords} -- One of the key requirements for the worm to be able
to spread was to be able to break the password of its current host. This was
done by exploiting the fact that encrypted user passwords were stored in a
publically readable file. However, in his technical report
\cite{techreportBySpaf} Spafford says that the passwords were not (he
means in effect) encrypted as a block of zero bits were repeatedly encrypted
using the user password and the result thus obtained was stored in the
publically readable file. Interestingly Morris had done a case study
\cite{morrisPasswordSec} on this before the attack. So, breaking passwords was
easy by guessing a list of passwords, then encrypting them and comparing the
output with the stored value.   
\end{enumerate}

\subsection{Innocence or not?}
\label{sec:lphm:inno} This subsection talks about the aspect that whether the
intent was malicious or not.  \paragraph{} Though the Cornell Commission
\cite{cornellCommission} states that the worm did not harm any user data or system
files, it did make infected systems slow. However, the commission also does not fail to
mention that "given Morris's evident knowledge of systems and networks, he knew
or clearly should have known that such a consequence was certain, given the
design of the worm".\\
The commission report also states
that Morris made only minimal efforts to halt the worm once it started
spreading and also accuses him of not informing any responsible authority
about it. However, according to these press
articles, Morris did try to talk to a friend, Graham, and Harvard University
who informed Andy Sudduth \cite{washingtonPost, heraldjournal}. Morris did suggest Sudduth some
steps to protect Harvard computers from the worm. Sudduth also says that after
some time Morris again called him realizing that he had made a 'colossal'
mistake, asking him to publicly publish an anonymous apology with instructions
on how to fix things. However, Sudduth was also the first witness for the
defense in the law-suit against the worm and, in response to a question by
prosecutor Mark D. Rasch, he said \cite{heraldjournal} that "He (Morris) wanted
that I should keep it quiet(a major security bug in a file transfer program in
the BSD UNIX), yes". This fact gives a reason to doubt that Morris might have
had notorious intents.\\ 
With above contrasting information, it is not clear 
whether Morris was only performing an innocent educational experiment with
no malicious intent. However, one thing is clear even if it was an
innocent act the Internet had grown to such an extent that even innocence can
cause great damage.  \paragraph{} Another interesting fact is that Eric Allman, developer of the sendmail and delivermail, in a personal communication to Donn Seeley said, "The trap door resulted from two
distinct 'features' that, although innocent by themselves, were deadly when
combined(kind of like binary nerve gas)" \cite{techreportByDonn}. Exploit of \textit{sendmail} is
another example of how innocent mistakes can be harmful.

%% file: wormAndPatches.tex
\section{Overview of the Worm and early actions} \label{sec:overviewAndPatches}
This section provides high-level overview of working of the worm \textit{i.e.}
answers the following question, "What exactly did the worm do that led it to
cause an epidemic?". This section will
also briefly discuss some of the early patches released to target this problem.
\paragraph{} The worm can be considered as divided into two major parts a
bootstrap(vector) program and a main program. The vector program is a 99-line C
program and the main program is a large relocatable object file that is
compatible with VAX and Sun-3 systems \cite{techreportByDonn}. The bootstrap
program is included in the appendix in \cite{techreportBySpaf}.
\subsection{Working of the worm} \label{sec:overviewAndPatches:working}
\subsubsection{How does the worm spread?}
Once the worm is established on a machine it then tries to locate a host and, most importantly, target accounts on the host to infect new machines which it then exploits via one of the loop holes discussed in \S\ref{sec:lphm} to pass a copy of the worm to the remote machine. 
The worm tries to obtain the address of potential target hosts by reading various system tables like \textit{/etc/hosts.equiv} and \textit{/.rhosts} and user files like \textit{.forward} and \textit{.rhosts}, ordered in such a way such that it reads files having the name of local machines first. It might be doing this as local machines are more likely to give access without authentication.
For a fixed address the worm can try to penetrate in one of the following ways:
\begin{itemize}
\item Exploiting the bug in the \textit{finger} server which lets the worm download its code instead of a \textit{finger} request and then tricking the server to execute it. 
\item Using the bug in the debugging code of the \textit{sendmail} SMTP mail service.
\item Guessing passwords and then using \textit{rsh} and \textit{rexec}
\end{itemize}
\subsubsection{Why was performance of the infected machines degraded?}
Infected machines slowed down because of uncontrollable replication of the worm because the worm was only using the main memory (see \S\ref{sec:overviewAndPatches:working:hide}) for its entire processing, this lead to memory clogging \cite{cornellCommission} and resulted in the machines slowing down.
\subsubsection{How does it try to hide itself?} \label{sec:overviewAndPatches:working:hide}
Following steps summarize some
of the masquerading steps taken by the worm described in
\cite{techreportByDonn} 
\begin{itemize} 
\item On startup, the worm used to delete its argument list and set the very
first argument as \textit{sh} in an attempt to pretend to be a shell command
interpreter.  
\item Used to \textit{fork} itself so that it doesn't have to stay with the same
process \textit{i.d.} for very long.
\item It read all the files which are part of the worm program into memory and
deletes them so that no evidence is left 
\item Turns off \textit{core-dump} generation, so that if the worm program crashes no dump files, leaving evidences behind are generated. This also helps in preventing analysis of the
worm.
\item While loading the worm file most of the non-essential symbol table
entries were deleted to ensure that even if the worm file is caught before deleting, it will be harder to guess what the routines
are doing.
\end{itemize} 
\subsubsection{What does it not do?} The program was a worm, and not a virus
(see \S\ref{sec:appendix:worm} and \S\ref{sec:appendix:virus} for difference in a worm and a virus). It did not
attempt to modify any other program or files on the system. It also, didn't
delete any system files \cite{techreportByDonn, cornellCommission}, and in-fact 
did not attempt to incapacitate the system by deleting any of the already existing
files, it only deleted files created by itself. It, also neither installed 
\textit{Trojan Horses} nor transmitted decrypted passwords anywhere. 
It didn't try to get \textit{superuser} privileges.

\subsection{Early Actions and Patches} Scientist and engineers at major institutions like MIT, NASA, Purdue
University, UC Berkeley, and many others started realizing that something was
wrong late night on Nov. 2, 1988, or early morning Nov. 3, 1988, and started
taking immediate action. The first formal public posting about the virus was
sent by Peter Yee of NASA Ames at 2:28 am, on Nov. 2, 1988, via the mailing list
\textquote{tcp-ip@sri-nic.arpa} \cite{mitRochlis}. Reports gave a detailed timeline of the
major events and actions from the assumed "beginning" of the worm until its full
decompiled code was installed at Berkeley \cite{techreportByDonn, mitRochlis}.

\paragraph{}On Thursday, Nov. 3 morning at 5:58 am several patches to
fix the worm were released by Keith Bostic of UC Berkeley
\cite{bosticWikipediapage}, one of the key people in the history of BSD UNIX,
via the \textit{tcp-ip} mailing list which was also forwarded by several
others\cite{techreportByDonn, mitRochlis}. E.H. Spafford
\cite{spafWikipediapage} from Purdue University analyzed the worm and released some patches
in his detailed technical report on the worm\cite{techreportBySpaf}. Patches
for \textit{fingerd} and \textit{sendmail} are discussed in detail in
this section. 
\paragraph{} Nov. 8, 1988, on the Tuesday following the event, the National Computer
Security Center(NCSC) called a meeting of scientists,
officers, faculty members from institutions including the National Institute of
Science and Technology, the Defence Communication Agency, the DARPA, the
Department of Energy, the Ballistics Research Laboratory, the Lawrence
Livermore National Laboratory, the CIA, the UC Berkeley, the MIT, SRI
International, the FBI, and various other stake holders. The three fourth of
the day was spent in analyzing the event from the perspective of all the
participants and the remaining time was used to discuss learnings from the
event and what actions to be taken \cite{mitRochlis}. Some of the actions taken by NCSC are
discussed in section \S\ref{sec:changes:system}.

\subsubsection{Patches for \textit{sendmail}} On the Thursday following the
attack, Keith Bostic sent following two fixes or suggestions for 
\textit{sendmail}: 
\begin{enumerate}
\item At 5:58 a.m. on the list \textit{tcp-ip@sri-nic.arpa} which provided the
\textit{compile without the debug command} fix to \textit{sendmail}
\cite{mitRochlis}. This posting also suggested to rename the UNIX C
compiler(cc) and loader(ld), this worked as the worm needed the path to them to
compile itself and helped in protecting against the \textit{non-sendmail} attack.
\item At 11:12 a.m. on the list \textit{comp.4bsd.ucb-fixes}. This fix
suggested to use \textit{0xff} instead of \textit{0x00} in the binary patch to
the \textit{sendmail}. This was needed to support the previous patch. The
previous patch was effective however would have fallen to \textit{debug} mode
if an empty command line was sent. He also asked to look for string
\textquote{debug} in the \textit{sendmail binary} using the UNIX
\textit{strings command} and mentioned that if there is no presence of the
string then the version is definitely safe.
\end{enumerate}
The patch for \textit{sendmail} can be found in the appendix of this
\cite{techreportBySpaf} report.\\

\subsubsection{Patches for \textit{fingerd}} On the Thursday night following
the attack at 10:18 p.m., Bostic sent out a fix for \textit{fingerd}. Overall,
this was the third fix posted by him. This fix had new source code for
\textit{fingerd} which used \textit{fgets} instead of \textit{gets} and instead
of \textit{return} used \textit{exit}. This bug-fix post also had another
version of \textit{sendmail} which totally removed the \textit{debug} command.
See \S\ref{sec:appendix:getsNoFix} to know more about why \textit{gets} cannot
be fixed.  A patch for \textit{fingerd} can be found in the appendix of this
\cite{techreportBySpaf} report.\\

On Friday, following the attack at 5:05 p.m., Bostic released his fourth
bug-fix. This fix was different than the previous ones, It was a fix to the
worm \cite{mitRochlis}.

%% file: lessonsLearnt.tex
\section{Lessons Learnt} \label{sec:learning} Use of worm-like capability was not new
in this case researchers have tried it to enable automated operating system
patches across multiple networks\cite{njvcBlog}. This incident showed that, with technology
advancement, unintended or undesirable consequences can follow. 
Being the first of its kind this event gave many lessons to scientists, 
engineers, technical agencies as well as general public. If we look at
the broader timescale around this event, this was the
time when the use of the Internet was growing rapidly outside of research.
Press reporting of the event (like this \cite{heraldjournal} and many others) 
made the general public (non-research community) aware of computer 
networking and, most importantly, it made many people aware of the fact that 
malicious computer programs can be written \cite{lessonsIntelNewsroom} and 
also made the community aware of computer and network security issues and a 
widespread concern grew out among people that whether the network no which 
many essential things like transportation, commerce, and high-risk services like national defense and space missions rely upon is secure. 
The shocking slowdown of the system and the awareness created made government agencies, universities, and all other stakeholders analyze the incident carefully and think about computer and network security from
all perspectives, and figure out learnings from the episode for future safety \cite{NCSCpostMortem}, \cite{mitRochlis} etc. Some of the views put forward by different people were contrasting. However, based on my readings
I found that there was more or less a general consensus among all the stakeholders regarding the following points.\\
\textbf{Diversifying the options} - If we make an analogy with biological fact, biological genetic diversity makes species more robust. Having diversity in computer programs will make the network more robust as an attack exploiting flaws of any particular software will have its effect limited only to machines running or using that software, and will prevent it from bringing the whole network down.\\
\textbf{Least Privilege} - The basic security principle of \textit{least privilege} says that any entity should only be given just enough privilege to carry out the work they are supposed to. There was a consensus among people that the principle of \textit{least privilege} should not be ignored for computer security. Ignoring this might lead to disasters.\\
\textbf{Defense Mechanism} - Defense mechanisms should be installed at end-hosts, in this particular event the network performed well and faults were in the application programs.\\
\textbf{Need of a Response and coordinating team} - Just like for any other critical situation there should be an emergency response team for responding to computer security threats \\
\textbf{No limited connectivity} - Researchers outright denied suggestions of limiting the connectivity and agreed to the argument that "\textit{the cure shouldn't be worse than the disease}" \cite{mitRochlis}. Limiting connectivity will negatively impact the progress of the research.\\

\subsection{Have we really learned? -- Status after 29 years} \label{sec:learning:really}
In that period, 1988 - 1990, for spreading the worm which slowed down some machines Morris was sentenced to 3 years of Probation and \$3276 to cover the cost of the probation, a fine of \$10,500 and 400 hours of community service. There was a sense of urgency among people -- many technologies and security-related(including non-computer security) agencies met within days of the event\cite{NCSCpostMortem}. Teams like \textit{CERT} were created and many other actions described in \S\ref{sec:changes} were taken, public awareness on network security spiked. Now, 29 years later, we are in a world where hackers are mostly known as someone who does money related fraud or steals digital information and uses it for some unintended purpose. The Morris worm has become a history lesson, a forerunner that put everyone on guard from a demon (However, if executed even now the worm will still need d\textbf{a}emons for carrying out its task! Pun intended).

As discussed earlier in this section, the event gave many immediate lessons but have we been able to keep up with learnings the forerunner gave us. Have we learned a hidden lesson, that if things can then they will go wrong? Have we been able to scale up the learnings and concern for computer and network security with growing use of digital devices and computer networks? In my opinion, the answer is yes and no. To me, it seems that in many cases we are on the "best effort" mode for computer and network security. 
Many core-technical companies do understand the importance of security and are working towards making systems and designs more secure. They have dedicated teams working towards product security \cite{googleSecurity}. 
System designers are taking care of security issues in advance while designing new systems or architecture. Security considerations in the design of ICN and NDN \cite{ICNSecuritySurvey,NDNBurke} are great examples. Anti-virus and anti-malware have a big market, and many products with varying prices, features and options available. They are also being widely used by people.

However, there are still many attacks happening and the number is increasing day by day. If we take a look at some of the top data breach attacks of 21$^{st}$ century \cite{topSecBreach21Century} more than 70\% of the victim companies are not core technical companies and are just using technology as a tool for their business. 
Attacks like \textit{Wanna Cry ransomeware} \cite{wannaCryWiki}, which are of much greater impact and severity as compared to the \textit{Morris Worm}, are occurring. In many cases, we are not even able to track the attacker, a conviction is out of the question.
Very few digital products(excluding those which are specifically designed for security) highlight security features in the product description.
The following could be the potential reasons behind the above problems that either leadership of the non-core tech companies and other regulatory authorities are not aware of or are not able to judge or willfully ignore to acknowledge the severity of risks involved in the computer security. Legal policies are not strong enough or detailed and precise enough to capture the computer security requirements \cite{digitalSecByHansen,cyberSecByMyriam}. Another potential reason could be that people are not able to correctly sense and measure the risk involved with the digital world as it is still very new. People do not use security features as one of the selection criteria for digital products.
Or maybe agencies still believe and agree with the statement that "It may be more expensive to prevent such attacks than it is to clean up after them" \cite{mitRochlis}.
When the worm incident happened most of the people using computers and networks were professionals and had significant knowledge of design and working of the system. However, today this not at all the case computers are just like any other equipment and many people with little, if any, knowledge about how system works are using it. There has been no significant step taken yet to educate such people. Schools and other public and private institutions do arrange time-to-time fire drills, and drills to handle other disaster events. However, I am not aware of or have heard of computer security drills. 
Even many Computer Science departments do not have a single mandatory course on computer security \cite{purdueCSBSRequirements}. However there are a large number of vacancies for cybersecurity experts, as per \cite{forbesCyberSecExpertsGap} every year in the United States., 40,000 jobs for information security analysts go unfilled, and institutions are struggling to fill 200,000 other cyber-security related positions.
I am not sure about lack of awareness but there definitely is lack of attention to computer security. Computer security is not getting the importance it should.



%% file: changes.tex
\section{What changed?} \label{sec:changes} Before the incident the Internet
was a closed community, and ethical and responsible behavior and good intent
were assumed. The worm episode changed people's behavior and attitude
towards computer security. Various actions were taken by different
organizations \cite{techreportByPeter, techreportByUSGeneralAccountingOffice},
government agencies, and universities, and of course, an event with such huge impact will definitely bring some changes in practice and perspective of
people. This section discusses these changes following the event.

\subsection{Changes in perspective} Post worm people had this sense that the Internet
is no longer a closed community. As published in an interview by Intel news room
\cite{lessonsIntelNewsroom} -- The day after the worm an awful lot of people
were shocked that such abuse could happen. This is evident from the
following statement in \cite{cornellCommission} "A community of scholars should
not have to build walls as high as the sky to protect reasonable expectation of
privacy, particularly when such walls will equally impede the free flow of
information." The statement clearly shows the disappointment of the community
on observing the breach of trust. This incident made a lot of people outside
the academic community aware of the fact that malicious software could be
written. From being a closed and trusted community of researchers the Internet
started becoming a large community, and it had to accept uncontrollable
sociopaths as it members. Computer science departments all across the world
started defining the appropriate and inappropriate usage of the Internet
resources.

\subsection{Changes in system} \label{sec:changes:system}
This subsection discusses more concrete and
administrative changes and other objective changes made in response to the 
episode. There were many administrative actions suggested and taken \cite{techreportByPeter,techreportByUSGeneralAccountingOffice,NCSCpostMortem}. Like
the formation of Computer Emergency Response Team(CERT) at Carnegie Mellon University with funding support from
DARPA. CERT was formed by organizing computer scientists with aim of isolating
such problems and preventing them from happening in the future. Though the
firewall technology already existed before the event, it saw a surge
 after the disturbing event \cite{fifteenYear}. Some people at DEC started
putting an effort into corporate network protection.\\ In a report to the Chairman of Subcommittee on
Telecommunications and Finance, Committee on Energy and Commerce, House of
Representatives US General Accounting Office recommended formation of an
interagency group including agencies funding research networks on the Internet,
under the coordination of President's Science Advisor, with following goals
(extracted from the report \cite{techreportByUSGeneralAccountingOffice}).

\begin{enumerate} 
\item provide Internet-wide security policy, direction and
coordination.  
\item support ongoing efforts to enhance Internet security.
\item obtain the involvement of Internet users, software vendors, technical
advisory groups, and federal agencies regarding security issues. 
\item become an integral part of the structure that emerges to manage the
National Research Network.  
\end{enumerate}

%% file: costAna.tex
\section{Cost Analysis: then and now} \label{sec:costAna}
This section subjectively discusses the direct or indirect costs incurred 
due to the worm and what it would cost for such an event in today's scenario.
With approximately only 6000 (10\% of total machines) \cite{cornellCommission}
machines being affected and the size of the network being only 60,000 machines, use of computers was esoteric 
and primarily for research, with high chances of the attacker not having 
criminal intent the losses were in millions of dollars. Experts estimated
that the per machine loss could be up to \$53,000 \cite{morrisCostFigures} and
another estimate in \cite{cornellCommission} said around 6000 machines 
were infected, so the total loss could be as high as \$318 million.\\
In contrast today there are 1) billions of digital devices in use and most of them
are connected to the Internet, 2) more than 51\% of total world population 3) Computers are being used
widely ranging trivial day to day work like managing traffic, listening to music, and education to high-risk tasks like medical treatment, national and internal security. A large number of
businesses entirely depend on computers and network
4) cyber-attacks being made with criminal intent \cite{wannaCryWiki,topSecBreach21Century}.
It is hard to imagine what will be the impact of an attack involving 10\% of total
machines on the Internet. It could simply break traffic in an entire city, state or maybe of whole nation. It could bring down several businesses or can even lead to large number of deaths if 
medical systems or security systems get into ambit of the attack.\\
According to cyber risk modeling from Cyence \cite{wannaCryLosses}, economic losses from a recent cyber attack, WannaCry ransomeware, \cite{wannaCryWiki}, could reach \$4 billion. The attack began on 12$^{th}$ May 2017 and impacted just 300,000 devices(less than 0.01\% of total digital devices) and 200,000 people (less than .003\%) of world population \cite{wannaCryWiki}. Even if extrapolate linearly the losses due to an attack involving 10\% of total devices on the network could be as high as \$4 trillion.


%% file: summary.tex
\section{Summary} \label{sec:summary}
This study compels me to agree to the statement (or its variations) by Prof.
Spafford that  \textquote{The only truly secure system is one that is powered
off, cast in a block of concrete and sealed in a lead-lined room with armed
guards — and even then I have my doubts} \cite{lessonsIntelNewsroom}. Any new
system starting with a closed group will have to deal with security issues as
it expands.  This is precisely what the Morris worm showed. An attack like this
was inevitable, had Morris not done it someone else would have. New upcoming
systems are now more cautious about security, ICN is an example. However, that
is not enough as discussed in \S\ref{sec:learning:really} we are not yet fully
ready to efficiently tackle cyber attacks like \cite{wannaCryWiki}. Other than
specialized training, there is pressing need to make general masses and policy
makers realize the importance of computer and Internet security. Demo cites{jajooSLearn, jajooPhilae, jajooGraviton, jajooSLearn, saathTechReport, phdthesis, philaeTechReport, slearnTechReport}

%% file: appendix.tex
\section{Appendix} \label{sec:appendix} 
\subsection{What is a computer worm?} \label{sec:appendix:worm} 
The concept of self-propagating worm program was first described by John Brunner
in his fictional novel \textit{The Shockwave Rider} \cite{brunnerNovel}. Worm in
the novel was used as a tool for taking revenge! \cite{wormWiki}. A computer worm
is a complete piece of program that replicates itself in order to spread to other
machines. Certainly, other than the very weak claim made in \cite{spafFailure} that warm
might have been written to take subconscious revenge on his father. I didn't find
any source claiming that Morris wrote the worm to take revenge. More about worms 
here \cite{wormTaxonomy}.

\subsection{What is a computer virus?} \label{sec:appendix:virus}
A computer virus is a malicious software. On execution, it replicates itself by 
"infecting" other programs \textit{i.e.} by inserting its own code into other computer programs. These infected programs can include data files or system
programs or even boot sector of the hard drive.

\subsection{Robert Morris's father was chief for computer security!} Robert
Tapan Morris's father, Robert Morris, was a computer scientist at Bell labs and
helped in designing Multics and Unix and later he became chief scientist at
National Computer Security Center, a division of the NSA. Some people claim
\cite{giacByLarry} that Junior Robert was trying to get away from his father's
image and have one of his own.

\subsection{\textit{gets} cannot be fixed.} \label{sec:appendix:getsNoFix} We
can say that functions implementing APIs like \textit{gets} will be
insecure-by-design.  Since these function will only get a pointer to a buffer(
char* ) and not the size of the buffer and since there is no implicit size
bound to a buffer in C, it will be impossible for the function to bound the
size of data being read.  This point is very clearly explained in the
\textit{linux man page} of \textit{gets} \cite{getsman}. \blockquote{Never use
gets(). Because it is impossible to tell without knowing the data in advance
how many characters gets() will read, and because gets() will continue to store
characters past the end of the buffer, it is extremely dangerous to use. It has
been used to break computer security. Use fgets() instead.}

%% file: morris.bib
@article{cornellCommission,
 author = {Eisenberg, T. and Gries, D. and Hartmanis, J. and Holcomb, D. and Lynn, M. S. and Santoro, T.},
 title = {The Cornell Commission: On Morris and the Worm},
 journal = {Commun. ACM},
 issue_date = {June 1989},
 volume = {32},
 number = {6},
 month = jun,
 year = {1989},
 issn = {0001-0782},
 pages = {706--709},
 numpages = {4},
 url = {http://doi.acm.org/10.1145/63526.63530},
 doi = {10.1145/63526.63530},
 acmid = {63530},
 publisher = {ACM},
 address = {New York, NY, USA},
}

@inproceedings{wormTaxonomy,
 author = {Weaver, Nicholas and Paxson, Vern and Staniford, Stuart and Cunningham, Robert},
 title = {A Taxonomy of Computer Worms},
 booktitle = {Proceedings of the 2003 ACM Workshop on Rapid Malcode},
 series = {WORM '03},
 year = {2003},
 isbn = {1-58113-785-0},
 location = {Washington, DC, USA},
 pages = {11--18},
 numpages = {8},
 url = {http://doi.acm.org/10.1145/948187.948190},
 doi = {10.1145/948187.948190},
 acmid = {948190},
 publisher = {ACM},
 address = {New York, NY, USA},
}

@article{fifteenYear,
  author={Orman, Hilarie},
  title={The Morris worm: A fifteen-year perspective},
  journal={IEEE Security \& Privacy},
  volume={99},
  number={5},
  pages={35--43},
  year={2003},
  publisher={IEEE},
}

@techreport{techreportBySpaf,
	author = {Eugene H. Spafford},
	title = {The Internet Worm Program: An analysis},
    howpublished = "\url{http://spaf.cerias.purdue.edu/tech-reps/823.pdf}",
	year = {1988},
	address = {Purdue University, West Lafayette, IN, USA},
    publisher = {ACM Computer Communication Review},
}

@techreport{jajooMwormResearchGate,
  title={A study on the Morris Worm},
  author={Jajoo, Akshay},
  year={2018},
howpublished = "\url{https://www.researchgate.net/profile/Akshay-Jajoo-3/publication/337499010_A_study_on_the_Morris_Worm/links/5ddc3f0a92851c1fedb1cfae/A-study-on-the-Morris-Worm.pdf}"
}

@article{jajooMwormPurdue,
  title={Term Paper on Morris Worm},
  author={Jajoo, Akshay},
  howpublished = "\url{https://www.cs.purdue.edu/homes/ajajoo/papers/morris-worm_term-paper.pdf}",
  year={2017}
}

@article{jajooMwormGithub,
  title={Term Paper on Morris Worm},
  author={Jajoo, Akshay},
  howpublished = "\url{https://github.com/ajajoo/morris-worm_term-paper/blob/master/morris-worm_term-paper.pdf}",
  year={2017}
}

@article{mitRochlis,
 author = {Rochlis, Jon A. and Eichin, Mark W.},
 title = {With Microscope and Tweezers: The Worm from MIT's Perspective},
 journal = {Commun. ACM},
 issue_date = {June 1989},
 volume = {32},
 number = {6},
 month = jun,
 year = {1989},
 issn = {0001-0782},
 pages = {689--698},
 numpages = {10},
 url = {http://doi.acm.org/10.1145/63526.63528},
 doi = {10.1145/63526.63528},
 acmid = {63528},
 publisher = {ACM},
 address = {New York, NY, USA},
}

@article{morrisPasswordSec,
 author = {Morris, Robert and Thompson, Ken},
 title = {Password Security: A Case History},
 journal = {Commun. ACM},
 issue_date = {Nov. 1979},
 volume = {22},
 number = {11},
 month = nov,
 year = {1979},
 issn = {0001-0782},
 pages = {594--597},
 numpages = {4},
 url = {http://doi.acm.org/10.1145/359168.359172},
 doi = {10.1145/359168.359172},
 acmid = {359172},
 publisher = {ACM},
 address = {New York, NY, USA},
}

@inproceedings{spafFailure,
  title={A Failure to Learn from the Past},
  author={Spafford, Eugene H},
  booktitle={Computer Security Applications Conference, 2003. Proceedings. 19th Annual},
  pages={217--231},
  year={2003},
  organization={IEEE}
}

@article{spafAftermath,
 author = {Spafford, E. H.},
 title = {Crisis and Aftermath},
 journal = {Commun. ACM},
 issue_date = {June 1989},
 volume = {32},
 number = {6},
 month = jun,
 year = {1989},
 issn = {0001-0782},
 pages = {678--687},
 numpages = {10},
 url = {http://doi.acm.org/10.1145/63526.63527},
 doi = {10.1145/63526.63527},
 acmid = {63527},
 publisher = {ACM},
 address = {New York, NY, USA},
}

@article{cyberSecByMyriam,
author = {Dunn Cavelty, Myriam},
title = {From Cyber-Bombs to Political Fallout: Threat Representations with an Impact in the Cyber-Security Discourse},
journal = {International Studies Review},
volume = {15},
number = {1},
pages = {105-122},
year = {2013},
doi = {10.1111/misr.12023},
URL = { + http://dx.doi.org/10.1111/misr.12023},
eprint = {/oup/backfile/content_public/journal/isr/15/1/10.1111/misr.12023/2/15-1-105.pdf}
}

@article{digitalSecByHansen,
  title={Digital disaster, cyber security, and the Copenhagen School},
  author={Hansen, Lene and Nissenbaum, Helen},
  journal={International studies quarterly},
  volume={53},
  number={4},
  pages={1155--1175},
  year={2009},
  publisher={Wiley Online Library}
}

@inproceedings{NDNBurke,
  title={Securing instrumented environments over content-centric networking: the case of lighting control and NDN},
  author={Burke, Jeff and Gasti, Paolo and Nathan, Naveen and Tsudik, Gene},
  booktitle={Computer Communications Workshops (INFOCOM WKSHPS), 2013 IEEE Conference on},
  pages={394--398},
  year={2013},
  organization={IEEE}
}

@article{ICNSecuritySurvey,
  title={Security, privacy, and access control in information-centric networking: A survey},
  author={Tourani, Reza and Misra, Satyajayant and Mick, Travis and Panwar, Gaurav},
  journal={IEEE Communications Surveys \& Tutorials},
  year={2017},
  publisher={IEEE}
}

@techreport{techreportByPeter,
	author = {Peter J. Denning},
	title = {The Internet Worm},
    howpublished = "\url{https://ntrs.nasa.gov/archive/nasa/casi.ntrs.nasa.gov/19900014594.pdf}",
    publisher = {RIACS Technical Report TR-89.3},
	year = {1989},
    address = {Research Institute for Advanced Computer Science, NASA Ames Research Center},
}

@techreport{techreportByUSGeneralAccountingOffice,
	author = {Ralph V. Carlone},
	title = {Computer Security: Virus highlights Need for Improves Internet Management},
    howpublished = "\url{https://nsarchive2.gwu.edu/NSAEBB/NSAEBB424/docs/Cyber-005.pdf}",
	publisher = {United States General Accounting Office},
	year = {1989},
    address = {United States General Accounting Office, Washington, D.C. 20548},
}

@techreport{techreportByDonn,
	author = {Donn Seeley},
    title = {A tour of the worm},
    publisher = {School of Computing, University of Utah},
    year = {1989},
    howpublished = "\url{https://collections.lib.utah.edu/details?id=702918}"
}

@misc{giacByLarry,
  author = {Larry Boettger},
  title = {{The Morris Worm: How it Affected Computer Security and Lessons Learned by it}},
  howpublished = "\url{https://www.giac.org/paper/gsec/405/morris-worm-affected-computer-security-lessons-learned/100954}",
  year = {2000}, 
}

@misc{tourByDonn,
  author = {Donn Seeley},
  title = {{A tour of the worm}},
  howpublished = "\url{http://www.cs.unc.edu/~jeffay/courses/nidsS05/attacks/seely-RTMworm-89.html}",
}

@misc{njvcBlog,
	title = {Morris Worm Turned on Cyber Security 25 years Ago},
    howpublished = "\url{http://www.njvc.com/blogs/morris-worm-turned-cyber-security-25-years-ago}",
    publisher = {NJVC},
    year = {2013},
}

@misc{kasperSky,
    title = {Morris Worm Turns 25},
    howpublished = "\url{https://www.kaspersky.com/blog/morris-worm-turns-25/3065/}",
	publisher = {Kasper Sky Blog},
    year = {2003},
}

@misc{forbesCyberSecExpertsGap,
	title = {The Fast-Growing Job With A Huge Skills Gap: Cyber Security},
    author = {Jeff Kauflin},
    howpublished = "\url{https://www.forbes.com/sites/jeffkauflin/2017/03/16/the-fast-growing-job-with-a-huge-skills-gap-cyber-security}",
    publisher = {Forbes},
    month = Mar,
    year = {2017},
}

@misc{washingtonPost,
  author = {Timothy B. Lee},
  title = {{How a grad student trying to build the first botnet brought the internet to its knee}},
  howpublished = "\url{https://www.washingtonpost.com/news/the-switch/wp/2013/11/01/how-a-grad-student-trying-to-build-the-first-botnet-brought-the-internet-to-its-knees/}",
  year = {2000},
  publisher = {The Washington Post},
}

@misc{lessonsIntelNewsroom,
	title = {Lessons from the 1st major computer virus},
    howpublished = "\url{https://newsroom.intel.com/editorials/lessons-from-the-first-computer-virus-the-morris-worm/}",
    publisher = {Intel Newsroom},
    year = {2013},
}

@misc{wikipedia,
  title = {{Morris worm}},
  howpublished = "\url{https://en.wikipedia.org/wiki/Morris_worm}",
}

@misc{heraldjournal,
	title = {Herald-Journal},
    year = {1990},
    howpublished = "\url{http://media.syracuse.com/vintage/other/2016/01/20/Testifies\%20merged.pdf}",
}

@misc{topSecBreach21Century,
	title = {The 16 biggest data breaches of the 21st century},
    year = {2017},
    howpublished = "\url{https://www.csoonline.com/article/2130877/data-breach/the-16-biggest-data-breaches-of-the-21st-century.html}",
}

@misc{googleSecurity,
	title = {Google's Security Culture},
    author = {Google Cloud},
    howpublished = "\url{https://gsuite.google.com/learn-more/security/security-whitepaper/page-2.html}",
}

@misc{morrisCostFigures,
	title = {Computing's 11 Smartest Super-Viruses-And The Damage They Wrought},
    howpublished = "\url{https://www.fastcompany.com/3015224/computings-11-smartest-super-viruses-and-the-damage-they-wrought}",
    publisher = {Fast Company},
    year = {2013},
    month = Feb,
}

@misc{wannaCryLosses,
	title = {"WannaCry" ransomware attack losses could reach dollar 4 billion},
    author = {Jonathan Berr},
    howpublished = "\url{https://www.cbsnews.com/news/wannacry-ransomware-attacks-wannacry-virus-losses/}",
    publisher = {Money Watch, CBS News},
	year = {2017},
    month = May,
}

@misc{wormWiki,
	title = {Computer Worm History},
    publisher = {Wikipedia},
    howpublished = "\url{https://en.wikipedia.org/wiki/Computer_worm\#History}",
}

@misc{wannaCryWiki,
	title = {Wanna Cry ransomeware attack},
    publisher = {Wikipedia},
    howpublished = "\url{https://en.wikipedia.org/wiki/WannaCry_ransomware_attack}",
}

@misc{globalInternetUsage,
	title = {Global Internet Usage},
    publisher = {Wikipedia},
    howpublished = "\url{https://en.wikipedia.org/wiki/Global_Internet_usage}",
}

@misc{morrisWikipediapage,
	title = {Robert Morris Wikipedia page},
	howpublished = "\url{https://en.wikipedia.org/wiki/Robert_Tappan_Morris}",
}

@misc{bosticWikipediapage,
	title = {Keith Bostic Wikipedia page},
	howpublished = "\url{https://en.wikipedia.org/wiki/Keith_Bostic}",
}

@misc{spafWikipediapage,
	title = {Eugene Howard Spafford Wikipedia page},
	howpublished = "\url{https://en.wikipedia.org/wiki/Gene_Spafford}",
}

@misc{morrisHomepage,
	title = {Robert Morris Homepage},
	howpublished = "\url{https://pdos.csail.mit.edu/archive/rtm/}",
}

@book{brunnerNovel,
	title = {The Shockwave Rider},
    author = {John Brunner},
    year = {1975},
}

@misc{getsman,
	title = {gets(3) - Linux man page},
    publisher = {die.net},
    howpublished = "\url{https://linux.die.net/man/3/gets}",
}

@misc{NCSCpostMortem,
	title = {Proceedings of the virus post-mortem meeting. },
	publisher = {National Computer Security Center}, 
	address = {Ft. George Meade, MD},
    year = {1988},
    month = Nov,
}

@Misc{purdueCSBSRequirements,
	title = {Bachelor of Science Degree Requirements},
    publisher = {Department of Computer Science, Purdue University},
    howpublished = "\url{https://www.cs.purdue.edu/undergraduate/curriculum/bachelor.html}"
}
